\documentclass[conference]{IEEEtran}
\IEEEoverridecommandlockouts
\usepackage{cite}
\usepackage{amsmath,amssymb,amsfonts}
\usepackage{algorithmic}
\usepackage{graphicx}
\usepackage{textcomp}
\usepackage[normalem]{ulem}
\usepackage{xcolor}
\usepackage{epstopdf}
\usepackage{url}
\usepackage{bm}
\usepackage{bbm}
\usepackage{upgreek}
\usepackage[super]{nth}
\usepackage[margin=0.75in]{geometry}
\usepackage{cancel}

\newcommand{\Honull}{\mathcal{{D}}_0}
\newcommand{\Hoalt}{\mathcal{{D}}_1}

\def\BibTeX{{\rm B\kern-.05em{\sc i\kern-.025em b}\kern-.08em
    T\kern-.1667em\lower.7ex\hbox{E}\kern-.125emX}}
\begin{document}
\normalem

\title{Artificial Intelligence and Location Verification in Vehicular Networks\\
}

\title{{Artificial Intelligence and Location Verification in Vehicular Networks}}
\author{
\IEEEauthorblockN{Ullah Ihsan$^1$, Ziqing Wang$^1$, Robert Malaney$^1$, Andrew Dempster$^1$ and Shihao Yan$^2$}
\IEEEauthorblockA{$^1$School of Electrical Engineering  \& Telecommunications,\\
The University of New South Wales,\\
Sydney, NSW 2052, Australia\\
$^2$School of Engineering,\\
Macquarie University,\\
Sydney, NSW 2109, Australia}}

\maketitle

\begin{abstract}
Location information claimed by devices will play an ever-increasing role in future wireless networks such as 5G, the Internet of Things (IoT).
Against this background, the verification of such claimed location information will be an issue of growing importance. A formal information-theoretic Location Verification System (LVS) can address this issue to some extent, but such a system usually operates within the limits of  idealistic assumptions on  \emph{a-priori} information on the proportion of genuine users in the field.  In this work we address this critical limitation by using a Neural Network (NN) showing how such a NN based LVS is capable of efficiently functioning even when the proportion of genuine users is completely unknown \emph{a-priori}.  We demonstrate the improved performance of this new form of LVS based on Time of Arrival measurements  from multiple verifying base stations within the context of vehicular networks, quantifying how our NN-LVS outperforms the stand-alone information-theoretic LVS in a range of anticipated real-world conditions. We also show the efficient performance for the NN-LVS when the users' signals have added Non-Line-of-Site (NLoS) bias in them. This new LVS can be applied to a range of location-centric applications within the domain of the IoT.
\end{abstract}


\section{Introduction}

We are at the verge of new  wireless networks that aim to bring communication revolutions in homes, hospitals, education,  transportation, and other aspects of  society. Emerging Intelligent Transportation Systems (ITS) are particularly exciting due to their potential to save many lives. The success of these new technologies in general, and ITS in particular, find their roots in the true location information of the clients (devices, users, vehicles) involved. In many scenarios it is anticipated that clients can directly obtain their location information \cite{hofmann2012global,bulusu2000gps} through the Global Navigation Satellite System (GNSS). This location information is then usually provided by the clients  to other clients or to some central Processing Center (PC), for verification and other network functionality purposes. But what if a client provides incorrect information about its true location intentionally in an attempt to obtain some advantage over other users \cite{jaeger2012novel,yu2013detecting}? Such circumstances could also occur unintentionally due to  difficulty in recording the GNSS location information or faulty hardware issues.

The focus of the work in this paper is location verification in the Internet of Things (IoT), in general, and in ITS (a sub-application of IoT) in particular. The location verification framework proposed in this work is applicable to all IoT applications whose performance is related to a user's reported location. Within ITS, if a malicious user provides inaccurate location information and this goes unnoticed, the possible aftermath could range from  sub-optimal traffic routing all the way through to life-threatening collisions \cite{leinmuller2005influence,leinmuller2006greedy}. Verification of a user's reported location information is hence critical for successful operation in ITS \cite{leinmuller2005influence,papadimitratos2006securing,raya2007securing,papadimitratos2008secure,mauve2001survey}.
Due to this, Location Verification System (LVS) performance has been a research focus in ITS for well over a decade. Recently, several information-theoretic LVSs  have been devised \cite{yan2013location,rob2016location,yan2016location}. These LVSs operate under a set of well-defined rules and conditions. Additionally, they have limitations in addressing various anomalies since they usually assume idealized channel conditions \cite{yan2013location}. As such, information-theoretic LVSs usually possess performance limitations in  real-world situations. One of the most important of these  limitations is the \emph{a-priori} lack of knowledge on the proportion (fraction) of vehicles in the field that will be malicious (alternatively, the fraction that will be genuine).

Neural Networks (NNs) have recently brought breakthroughs into many aspects of modern society. Web mining\cite{abadi2016tensorflow}, content filtering\cite{wan2014deep}, image recognition\cite{krizhevsky2012imagenet,szegedy2015going}, speech processing\cite{hinton2012deep}, language identification \cite{saad}, speaker verification\cite{matvejka2016analysis}, object detection \cite{vaillant1994original,lawrence1997face}, advanced genomics\cite{libbrecht2015machine}, and drug discovery\cite{ramsundar2015massively} are just a few of the fields impacted. NNs also lay the foundation for many aspects of the self-driving car paradigm\cite{guizzo2011google}. Many of these breakthroughs are achieved through the development of new NN algorithms\cite{lecun2015deep}.

We consider an LVS based on Time of Arrival (ToA) measurements \cite{yan2014timing} under the influence of Non-Line-of-Sight (NLoS) biases. The novel contributions in this work are summarized below.
 \begin{itemize}
 \item We show that the NN-LVS proposed in this work outperforms an information-theoretic LVS\cite{yan2014timing} when the ToA of the users' signals have added NLoS bias in them.
 \item We also show that unlike the information-theoretic LVS which assumes an \emph{a-priori} knowledge about the proportion of malicious  vehicles in the field, the NN-LVS works satisfactorily in the complete absence of this knowledge.
 \end{itemize}

Recent advancements in digital signal processing and hardware design now provide us with very accurate physical-layer timing information for wireless networks\cite{steffes2011determining,martin2013software}. These developments provide us with the clock synchronization that enables the LVS we study here. As such, we suggest our new NN-LVS can offer a viable and  pragmatic solution to the important task of location verification for many IoT applications under real-world conditions and uncertainties.

The remainder of this paper is organized as follows. Section \ref{SM} describes the system model. Section \ref{PA} details the performance analysis using information theory and neural network techniques. Section \ref{NR} provides the numerical results, and Section \ref{FW} highlights future prospects. Section \ref{conclusions} concludes the paper.

\section{System Model}\label{SM}

We outline the system model and assumptions considered for our framework:
\begin{enumerate}
\item {The framework consists of $N$  trusted Base Stations (BSs) as verifiers with publicly known locations that are assumed to be in the range of the prover (the vehicle whose claimed location is to be authenticated). The location of the \textit{i}-th BS is $\bm{X}_i=[x_i,y_i]$ where $i=1,2,...,N$.}
\item {The true location from a genuine or malicious vehicle (the prover) is denoted by $\bm{X}_t=[x_t,y_t]$ and is assumed to possess  zero localization error.}
\item {We refer to the announced (reported) location from a legitimate or malicious vehicle as \emph{claimed location} and denote it by $\bm{X}_c=[x_c,y_c]$. For a legitimate vehicle, the claimed location is exactly the same as its true location. On the other hand, a malicious vehicle spoofs its true location to the BSs (to potentially obtain an advantage over other vehicles or to disrupt the system performance). The true location of a malicious vehicle is unknown to the wider network.}
\item {One of the $N$ BSs is chosen as the PC. Measurements from all BSs are collected at the PC before being processed into a  a binary decision related to a vehicle's claimed location.}
\item {Under the null  hypothesis $\mathcal{H}_o$, the framework assumes a vehicle to be legitimate, \emph{i.e.},}
\begin{align}
\mathcal{H}_o : X_c=X_t.
\end{align}
\item {Under the alternate hypothesis $\mathcal{H}_1$, the framework considers a vehicle to be malicious, \emph{i.e.},}
\begin{align}
\mathcal{H}_1 : X_c\neq X_t.
\end{align}
\end{enumerate}
Under $\mathcal{H}_o$, the ToA value measured by the \textit{i}-th BS from a legitimate vehicle is given by
\begin{align}
Y_i=U_i+X_i,\ \ \ \ \ \ \ \ \ \ \ \ i=1,2,\dots ,N,\
\end{align} where $X_i$, the BS's receiver thermal noise, is a zero-mean normal random variable with variance $\sigma^2_{T}$. $U_i$ is the ToA and is given by\\
\begin{align}\label{distance_time}
U_i=\frac{d_i^c}{c},
\end{align}
where $d_i^c$ is the Euclidian distance of \textit{i}-th BS to a legitimate vehicle's true location, and is given by
\begin{align*}
d_i^c=\sqrt{{(x_c-x_i)}^2+{(y_c-y_i)}^2},
\end{align*}
with $c$ as the speed of light.

We assume the measurements made by the \textit{N} BSs to be independent of each other. Under $\mathcal{H}_o$, they collectively form a vector $\boldsymbol{\mathrm{Y}}\boldsymbol{=}{[Y}_1,\ Y_2,\dots ,{Y_N]}^T$ . Vector $\boldsymbol{\mathrm{Y}}$ follows a multi-variate normal distribution given as
\begin{align}
\mathbf{Y}| \mathcal{H}_o \sim \mathcal{N} (\bm{U}, \bm{R}),
\end{align}
where $\bm{U} = {[U}_1,\ U_2,\dots ,{U_N]}^T$ is the mean vector under the null hypothesis, and $\bm{R} = \sigma_T^2\bm{I}_N$ is the covariance matrix.
\\\\
\noindent Under $\mathcal{H}_1$, a malicious vehicle claims to be at a location removed from its true location. In a real-world scenario, we can think of this as if the malicious vehicle pretends to be on the road when he actually is placed off the road in a street or in a building. The ToA value measured by the \textit{i}-th verifier from a malicious vehicle is given by
\begin{align}
Y_i=T_x+W_i+X_i,\ \ \ \ \ \ \ \ \ \ \ \ i=1,2,\dots ,N,\
\end{align}
where $T_x$ is a time bias potentially added by malicious vehicle which impacts the overall ToA value. $W_i$ is given by
\begin{align}
W_i=\frac{d^t_i}{c},
\end{align}
where $d^t_i$ is the Euclidian distance of \textit{i}-th BS to a malicious vehicle's true location, and is given by
\begin{align*}
d^t_i=\sqrt{{(x_t-x_i)}^2+{(y_t-y_i)}^2}.
\end{align*}
We assume the measurements made by \textit{N} BSs to be independent of each other. Under $\mathcal{H}_1$, they collectively form a vector $\boldsymbol{\mathrm{Y}}\boldsymbol{=}{[Y}_1,\ Y_2,\dots ,{Y_N]}^T$ . Vector $\boldsymbol{\mathrm{Y}}$ follows a multi-variate normal distribution given as
\begin{align}\label{likelihood_H1}
\mathbf{Y}| \mathcal{H}_1 \sim \mathcal{N} (\bm{W} + T_{x} \bm{1}, \bm{R}),
\end{align}
where $\bm{W} + T_x \bm{1} = [W_1+T_x, W_2+T_x, \dots, W_N+T_x]^T$ and $\bm{1}$  is a vector equal to the length of the number of BSs $N$ with all its elements set to 1. For later convenience we set $\bm{V}=\bm{W} + T_{x} \bm{1}$
and rewrite \eqref{likelihood_H1} as
\begin{align}
\mathbf{Y}| \mathcal{H}_1 \sim \mathcal{N} (\bm{V}, \bm{R}).
\end{align}

\section{Performance Analysis}\label{PA}

The outcome of an LVS is a binary result, \emph{i.e.}, legitimate or malicious. This is different from a localization system where the output is an estimated location. We measure the performance of our LVS using two methodologies; through information-theoretic analysis followed in \cite{yan2014timing} and, through the newly designed NN method which makes use of the machine-learning techniques. In both  cases, a Bayes average cost function is chosen as the performance metric for LVS in terms of the `Total Error'. The information-theoretic method is based on the \emph{a-priori} assumption that the proportion of malicious vehicles is known. Usually this is set to 0.5 in the absence of any other information. On the other hand, the NN-LVS calculates the Total Error irrespective of any such \emph{a-priori} assumption and it can function with any proportion of malicious vehicles in the field. The  Total Error is given by
\begin{align} \label{total_error}
\mathrm{\upxi} = p(\mathcal{H}_o)\upalpha \mathrm{\:+\:}p(\mathcal{H}_1)(1-\upbeta),
\end{align}
where $p(\mathcal{H}_o)$ and $p(\mathcal{H}_1)$ are the \emph{a priori} probabilities of occurrences of $\mathcal{H}_o$ (\emph{i.e.}, legitimate vehicle) and $\mathcal{H}_1$ (\emph{i.e.}, malicious vehicle), respectively, and are set equal, \emph{i.e.}, 0.5. $\upalpha$ represents the false positive rate (the rate of legitimate vehicles being detected incorrectly) and $\upbeta$ represents the detection rate (the rate of malicious vehicles being detected correctly). Equation \eqref{total_error} then takes the form
\begin{align}\label{totalerror}
\mathrm{\upxi} = 0.5\upalpha \mathrm{\:+\:0.5}\left(\mathrm{1-}\upbeta \right).
\end{align}

\subsection{LVS Performance Analysis Using Information Theory}

The Likelihood Ratio Test (LRT) is used for performance measurement of the LVS. It has been proven earlier that the LRT achieves the optimum detection results for a given false positive rate\cite{neyman1933ix}. This leads to the conclusion that the LRT minimizes the Total Error and maximizes the mutual information between input and output of LVS\cite{yan2014optimal}. We follow the  decision rule below for the LRT

\indent
\begin{align}\label{LRT}
\Lambda \left(\boldsymbol{Y}\right)\triangleq \frac{p(\boldsymbol{Y}|\mathcal{H}_1)}{p(\boldsymbol{Y}|\mathcal{H}_o)}\ \genfrac{}{}{0pt}{}{\genfrac{}{}{0pt}{}{\Hoalt}{\ge }}{\genfrac{}{}{0pt}{}{<}{\Honull}} \lambda,
\end{align}
where $\Lambda \left(\boldsymbol{Y}\right)$ is the likelihood ratio, $\lambda$ is the decision threshold, and $\Hoalt$ and $\Honull$ are the binary decision values (\emph{i.e.}, whether the vehicle is legitimate or malicious). Given the multi-variate normal form of the observations, \eqref{LRT} can be reformulated as

\indent
\begin{align} \label{LRT2}
\Lambda \left(\boldsymbol{Y}\right)=\frac{e^{\mathrm{-}\frac{1}{2}{\left(\boldsymbol{Y}-\boldsymbol{V}\right)}^{\mathrm{T}}\boldsymbol{\ }{\boldsymbol{R}}^{-1}\boldsymbol{\ }\left(\boldsymbol{Y}\boldsymbol{-}\boldsymbol{V}\right)}}{e^{\mathrm{-}\frac{1}{2}{\left(\boldsymbol{Y}-\boldsymbol{U}\right)}^{\mathrm{T}}\boldsymbol{\ }{\boldsymbol{R}}^{-1}\boldsymbol{\ }\left(\boldsymbol{Y}\boldsymbol{-}\boldsymbol{U}\right)}}\ \ \ \genfrac{}{}{0pt}{}{\genfrac{}{}{0pt}{}{\Hoalt}{\ge }}{\genfrac{}{}{0pt}{}{<}{\Honull}}\ \ \ \lambda.
\end{align}

\subsection{LVS Performance Analysis Using Neural Networks}

This section highlights the novel approach used to design a classification framework for the verification of a vehicle's claimed location through supervised machine-learning techniques. The framework uses a multi-layer feed-forward NN for the binary classification of a vehicle as either legitimate or malicious.

\begin{figure}[h]
\centering
\includegraphics[width=0.48\textwidth]{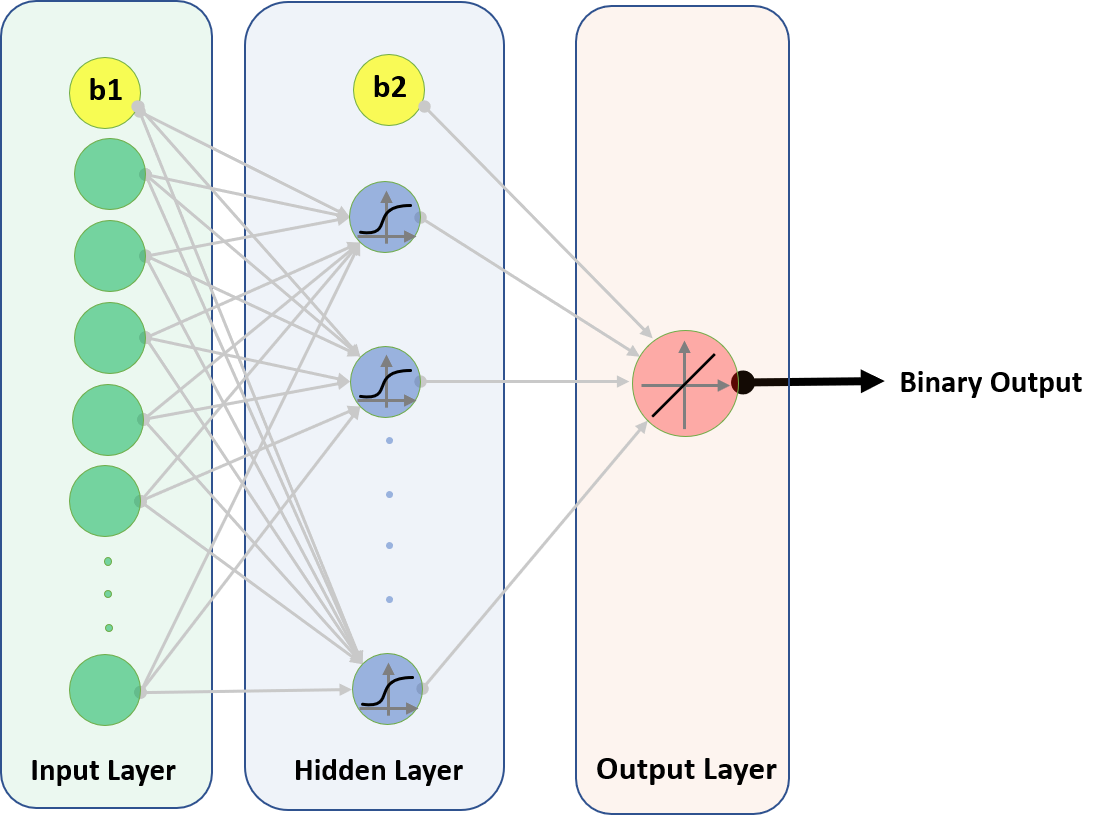}
\caption{The architecture of the Neural Network (NN) used for Location Verification in this work. This architecture arose from many trials of biased timings using different architectures with varying numbers of layers.\label{Fig1}}
\end{figure}

For uniformity, the framework considers the same inputs as considered for the information-theoretic method. These inputs include 
the ToA of the signal based on the user's claimed location, and $\mathit{\mathbf{Y}}$ (the observation vector influenced by the thermal noise $X_i$). Based on a series of trials with changing architectures for the NN-LVS, we finalised a framework that has an input vector, a hidden layer (with 10 neurons), and a binary output layer as shown in Fig.~\ref{Fig1}.
The NN LVS achieved optimum performance through the use of a Hyperbolic tangent sigmoid, and linear transfer functions in the hidden, and output layers, respectively.

\section{Numerical Results}\label{NR}

We now present some numerical results based on our analysis from the information-theoretic and NN-based LVS. In carrying out these simulations, BSs are located in a 1000 meters by 500 meters area at fixed publicly known locations. This area closely resembles a small district of a city and corresponds to the context of ITS where the BSs are trusted verifiers located on the roadside or in the nearby parking lots. The claimant vehicle (the prover whose location is yet to be verified) resides in a 500 X 500 meters area in between the BSs. In order to simulate the attacking scenario and thus study the performance of both the LVS, we assume there are two claimants that are within the communication range, namely a legitimate vehicle which is reporting its true location to the BSs and a malicious vehicle which is performing the location-spoofing attack. Both the vehicles can overhear the communication between the BSs and thus both acquire the locations of the BSs. If malicious, the vehicle can also overhear the communication between legitimate vehicles and the BSs so that it can forge its claimed location to that of the legitimate vehicle's true location.

The malicious vehicle sets its true location at a far-off point so that its transmitted signal (with the appropriate timing offset) has equal ToA at all the BSs (in the limit of the true location of a malicious vehicle being much greater that any other scale all NLoS biases at all BSs are the same). Under this approximation the mean ToA at the BSs is just the mean of the timings anticipated from a vehicle at the claimed location. The resultant alteration in ToA due to the receiver's thermal noise is extracted from a Gaussian random distribution with fixed standard deviation. The value of standard deviation considered in our simulation is set to 300 nanoseconds.

We use simulated ToA data in our numerical experiments. The claimed locations for genuine and malicious vehicles in equal proportion are generated randomly in the specified area. The ToA from the claimed locations at the 4 BSs is calculated using equation \eqref{distance_time}. The receivers in the BSs are under the influence of independent thermal noise $X_i$ and thus the ToA measurements they make have a certain degree of variation. We extract this variation (in nanoseconds) from a Gaussian random function that has a fixed standard deviation. The area around the vehicles is non LoS and therefore their transmitted signal cannot reach the BSs directly and hence their ToA have an additional NLoS bias $\phi_i$  in them. To mimic  reality, we extract $\phi_i$ from an exponential distribution as given below
\begin{align}\label{exponential_bias}
f(\phi_i) = \rho_i e^{-\rho_i \phi_i},
\end{align}
where $\rho_i$ is the scale parameter.

For the information-theoretic LVS, we calculate the Total Error, the false positive rate, and the detection rate using equations \eqref{totalerror} and \eqref{LRT2}, respectively.

The data considered for the information-theoretic LVS analysis is used to also train the NN-LVS. We call this data  the training data.\footnote{By training data we mean ToA data received from vehicles who we know \emph{a priori} to be legitimate or malicious. Use of such data in order to set the neural network parameters, prior to its use on `unlabeled' data (\emph{i.e.}, data from vehicles who we do not know \emph{a priori } to be legitimate or malicious), is known as the training phase.} In the training phase, we feed the NN-LVS with random vehicles data at a speed of one vehicle data per second. During each second the NN-LVS is trained with the available training data. The backpropagation algorithm has a set of internal parameters to terminate the training phase for the NN-LVS. We observe that in most of the cases the 'maximum validation failures'; which is the  maximum number of iterations in a row during which the NN-LVS's performance fails to improve or remains the same, terminates the training phase. We set this parameter to 6. The weights and biases are considered as optimised once the training phase has concluded. The NN-LVS afterwards can be used to classify a vehicle as genuine or malicious in the test data\footnote{The test data is simulated under a different realization with same settings as training data. Further, the test data has no labels.}.

The NN-LVS is trained during the \nth{1} second with an input training data from a single random vehicle. At the end of \nth{1} second, we subject the NN-LVS (with its weights and biases optimized) to calculate a Total Error for the test data. In the \nth{2} second, we add another random vehicle training data to the previously available single vehicle training data. The combined data forms a new training data set which is used to retrain the NN-LVS (from \nth{1} second). After a re-training, the NN-LVS is used to calculate a new Total Error for the test data. We add yet another random vehicle training data to the previously available training data in the \nth{3} second and use the updated data set to once again train the NN-LVS. At the end of the third second a revised Total Error is calculated for the test data. This process of updating the training data set, retraining the NN-LVS and recalculating a new Total Error for the test data continues in the following seconds. The Total Error keeps on decreasing with the passage of time.

\begin{figure}[t!]
\includegraphics[width=0.50\textwidth]{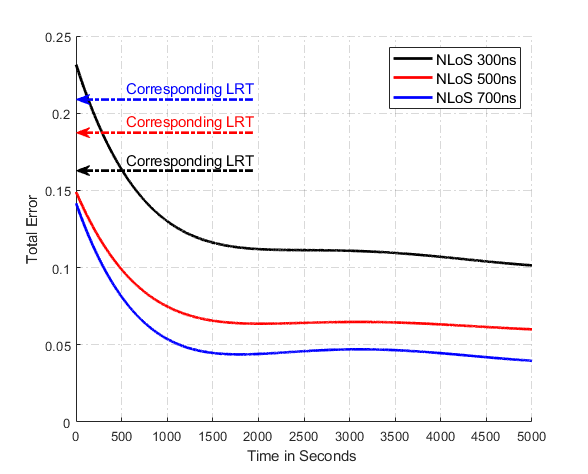}
\caption{Total Error performance of the NN-LVS with 4 BSs as it is being trained under changing NLoS bias conditions. The training and the testing data has genuine and malicious
vehicles in equal proportions. $X_i$ (extracted from a Gaussian random distribution) and NLoS bias (extracted from an exponential distribution and shown by different colour of the curves) both have a fixed standard deviation of 300 nanoseconds. The NN-LVS indicates an improved performance with a Total Error of 0.10 (NLoS 300ns), 0.06 (NLoS 500ns) and 0.04 (NLoS 700ns) as compared to the LRT method of \cite{yan2014timing} which gives a Total Error of 0.16 (NLoS 300ns), 0.18 (NLoS 500ns) and 0.21 (NLoS 700ns). Higher NLoS leads to easier discrimination between a genuine vehicle and a malicious one placed far from the BSs.\label{fig_main}}
\end{figure}

In Fig.~\ref{fig_main} we initially calculate the Total Error for a data set that has genuine and malicious vehicles in equal proportions (Po$=0.5$). The standard deviation for $X_i$ (extracted from a Gaussian distribution) is 300 nanoseconds while the standard deviation for NLoS bias (extracted from an exponential distribution) is indicated by the different curves. The number of BSs used is 4. The LRT (\emph{i.e.}, the Total Error arising from the information-theoretic LVS) corresponding to each NLoS curve is indicated by the dashed arrow lines. We can see that performance for the information-theoretic LVS deteriorates as the NLoS bias increases while the performance for the NN-LVS improves with an increase in the NLoS bias in the ToA data. It is clear that the NN-LVS is able to accommodate  the NLoS conditions significantly better than an information-theoretic LVS.

\begin{figure}[t!]
\includegraphics[width=0.50\textwidth]{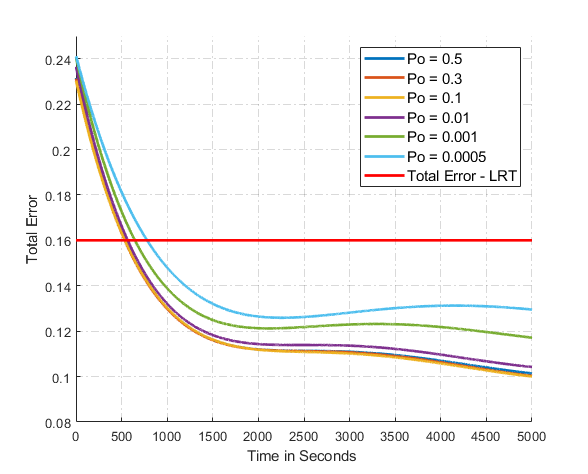}
\caption{Total Error performance of the NN-LVS with 4 BSs. The test data has different proportions for genuine and malicious vehicles as highlighted by the different colour of curves. $X_i$ is extracted from a Gaussian random distribution with a standard deviation of 300 nanoseconds. The NLoS bias is extracted from an exponential distribution with a fixed standard deviation of 300 nanoseconds. The red line shows the Total Error for the information-theoretic LVS (based on LRT method) for a data (realised under same settings of standard deviation for $X_i$ and NLoS bias) which has a Po equal to 0.5. We can see that the NN-LVS performs consistently with different Po(s) in the test data.\label{T300-N300-V4}}
\end{figure}

Next, we study the impact of changing Po (the proportion of malicious vehicles) over the performance of LVS. We train a NN-LVS through similar procedures as described earlier but change the proportion of malicious vehicles in the test data. In one of the experiments, we fix the standard deviation for $X_i$ and the NLoS bias both to 300 nanoseconds. The number of BSs are 4. We can see in Fig.~\ref{T300-N300-V4} that NN-LVS performs consistently even when Po is different in the test data.  We can see that NN-LVS performance is satisfactory even when the test data has 99.95\% genuine vehicles and 0.05\% malicious vehicles. The red line in the Fig.~\ref{T300-N300-V4} shows the Total Error for the information-theoretic LVS when the genuine and malicious vehicles are in equal proportions in the data (with no changes to $X_i$ and NLoS bias). Our study shows that unlike the information-theoretic LVS whose performance is conditioned to the \emph{a-priori} knowledge of Po, the NN-LVS's performance is largely independent of Po.

\begin{figure}[t!]
\includegraphics[width=0.50\textwidth]{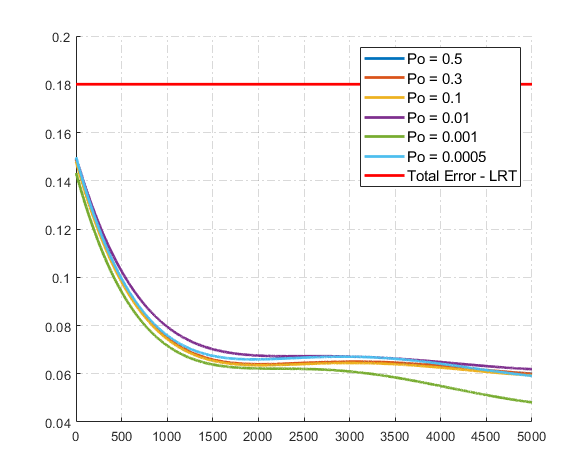}
\caption{Total Error performance of the NN-LVS as in Fig.~\ref{T300-N300-V4} except the standard deviation for NLoS bias now is 500 nanoseconds.}\label{T300-N500-V4}
\end{figure}

In Fig.~\ref{T300-N500-V4}, we change the standard deviation for NLoS bias to 500 nanoseconds. $X_i$ still is extracted from a Gaussian random distribution with a standard deviation of 300 nanoseconds. We can observe that NN-LVS's performance is independent of the Po value.

\begin{figure}[t!]
\includegraphics[width=0.50\textwidth]{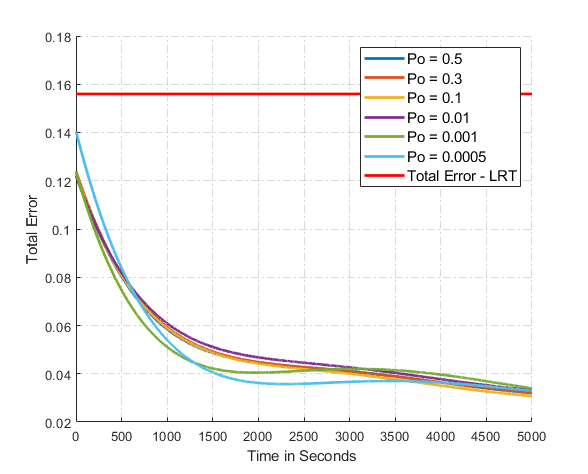}
\caption{Total Error performance of the NN-LVS as in Fig.~\ref{T300-N500-V4} except the number of BSs now are 6.}\label{T300-N500-V6}
\end{figure}

In Fig.~\ref{T300-N500-V6}, we change number of BSs to 6 while the standard deviation for NLoS bias and $X_i$ are kept the same as in Fig.~\ref{T300-N500-V4}. Still we can see a promising performance from the NN-LVS under different Po(s).

In another work, we also analyzed the performance of the NN-LVS proposed here by subjecting it to a real-world experimental situation where measured Received Signal Strength (RSS) inputs were used instead of simulated ToA values\cite{ihsan2019machine}. The RSS  from the vehicle was measured at 3 BSs. The performance results for the NN-LVS using input RSS measurements were consistent with the claims of this work, showing our NN-LVS's adaptability to other metrics beyond ToA.

The NN-LVS framework proposed in this work can be applied to many sub-applications within IoT and 5G whose performance largely depends on the true and verified locations of users. Map services, smart parking, Massive MIMO, Enhanced beamforming, coverage enrichment, and interference mitigation are just a few of the applications which can benefit from a NN-LVS. The same framework can also assist in defence-related operations where location and its verification is of prime importance.

\section{Future Work}\label{FW}

We aim to add more prominent features related to the channel environment so as to further investigate gains achieved by NN-based  LVSs. These additional features will better describe the real-world channels. We also plan to compliment ToA systems by adding RSS and Angle of Arrival (AoA) measurements thus enhancing the reliability of NN-based LVSs. A combination of ToA, RSS and AoA will make the NN LVS even more reliable and efficient in its location verification. We also plan to extend the NN framework to more complex channel fading models such as Rician fading channels. Estimating channel parameters through machine-learning will also result in an extended performance for the NN LVS. In this case the neural network architecture will need to be extended so as to accommodate the additional unknowns that must be learned.

\section{Conclusion}\label{conclusions}

Information-theoretic LVS frameworks, due to their operating limitations, are not practical in many real-world scenarios. To address this gap, we have proposed  the use of a neural network approach to location verification. This approach is particulary useful when we consider that one of the key inputs to any LVS is knowledge on the proportion of vehicles anticipated to be malicious - an input usually unknown. Using simulated ToA data, we have shown how a NN-based LVS outperforms a state-of-the-art information-theoretic LVS. Unlike the information-theoretic LVS, the working of the NN-LVS is shown to be largely independent of the proportion of  malicious vehicles in the  area. Unknown channel conditions, such as NLoS bias effects, were also shown to be better accommodated  by the NN-LVS approach.

More salient real-world channel features, will be considered in our future work to further improve the overall system performance. This will help us develop a more robust state-of-the-art artificially intelligent LVS, an LVS which will be wholly practical in terms of its location verification performance in a wide range of future wireless networks beyond the ITS we have studied here. We believe the novel approach for enhancing the performance of real-world LVSs that we have developed here potentially forms the foundation for all future works in this important area.

\section*{Acknowledgment}

The authors acknowledge support by the University of New
South Wales, Australia, and Macquarie University, Australia.
Ullah Ihsan acknowledges financial support from the Australian Government through its Research Training Program.

\bibliographystyle{IEEEtran}
\bibliography{Globcomm}

\end{document}